\documentclass[aps,prd,eqsecnum]{revtex4}

\usepackage{epsf}

\usepackage{psfrag}

\begin{document}

\title{Quasinormal modes of Kerr black holes: The determination of the quasinormal frequencies with a new technique}
 
\author{M. Giammatteo}
\affiliation{Department of Mathematics,\\University of Newcastle upon Tyne, NE1 7RU U.K.}

\date{ February 2003}

\begin{abstract}
We compute the quasinormal frequencies of rotating black holes using the continued fraction method first proposed by Leaver. The main difference with former works, is that our results are obtained by a new numerical technique which avoids the use of two dimensional root-finding routines. The technique is applied to evaluate the angular eigenvalues of Teukolsky's angular equation. This method allows us to calculate both the slowly and the rapidly damped quasinormal frequencies with excellent accuracy. 
\end{abstract}

\maketitle
\vspace{0.7cm}
\section{Introduction}
Gravitational waves emitted by perturbed black holes are well known to be dominated, at late times, by quasinormal modes. They are characterised by complex frequencies with positive imaginary parts and represent therefore damped oscillations. The quasinormal frequencies of black holes in asymptotically flat spacetime are of great importance from the astrophysical point of view. They give us information about the main parameters of a black hole such as its mass and its angular momentum. It is for this reason that they have been extensively studied for more than forty years.

The pioneering work on this topic was done by Regge and Wheeler \cite{regge} , who first studied linear perturbations of static black holes. They derived a second order ordinary differential equation which describes scalar, electromagnetic and gravitational perturbations of Schwarzschild black holes. The problem of separating the wave equation for Kerr rotating black holes, was solved by Teukolsky \cite {teukolsky}, who managed to obtain a couple of independent partial differential equations for the radial and angular part of the perturbation. An extensive study of black hole perturbation can be found in \cite{chandra}.

Quasinormal modes were first found by Vishveshwara \cite{vishe} and by Press \cite{press} through numerical computations of the time evolution of the gravitational waves around the blach hole. The first method for calculating them numerically was proposed by Chandrasekhar and Detweiler \cite{detw}, and it was based on the direct numerical integration of the Regge-Wheeler equation. There have also been analytical attempts to obtain these frequencies and all the methods have also been applied to the Kerr and Reissner-Nordstrom black holes. However, these methods cannot provide very accurate values for the frequencies and break down for rapidly damped modes.

Leaver \cite{leaver}, has shown that quasinormal frequencies of both static and rotating black holes can be determined by a continued fraction method. There are several methods to compute the quasinormal frequencies for static black holes, but the continued fraction method is still the only one which can be generalised to the Kerr case, where beside the quasinormal frequencies we must evaluate the separation constants, that is the angular eigenvalues associated with the Teukolsky equations. An improvement of Leaver's results can be found in \cite{onozawa}. 

In rotating black hole perturbation it is, indeed, the simultaneous computation of the angular eigenvalues that causes most of the problems of inaccuracy of the related frequencies. An alternative approach to Leaver's is to use ordinary perturbation theory, as Sasaki did \cite{sasaki}, to solve the angular equation, but this method only works as far as we consider small values for the expansion parameter $aw$. In other words, for non zero values of $a$, this technique does not allow us to deal with high frequencies.

In this paper we follow Leaver's continued fraction method, but after obtaining an analytical solution for the radial and angular equations, we do not solve the related two continued fraction equations simultaneously as he did. Our approach consists, first of all, in the computation of the angular eigenvalues by a Pade' approximation of the true values, and then in the use of this approximation into the radial solution. 

Where previous results existed, our results are in close agreement and therefore provide independent confirmation. We also give new results for higher frequency modes and tabulate the Pade' approximations for the angular modes, which can be used in a wide range of applications concerning rotating systems. 

In what follows, we use geometric units, that is $c=G=1$.
\newpage
\section{Angular and radial continued fraction equations}

In the Boyer-Lindquist coordinates $(t, r, \theta, \phi)$, rescaling $t$ and $r$ such that $c=G=1$, the Kerr metric is 
\begin{equation}
ds^2=\left(1-\frac{2Mr}{\Sigma}\right)dt^2+\left(\frac{4Mar\sin^{2}\theta}{\Sigma}\right)dt d\phi-\frac{\Sigma}{\Delta}dr^2-\Sigma d\theta^2-\sin^2 \theta\left(r^2+a^2+\frac{2Ma^2 r\sin^2 \theta}{\Sigma}\right)d\phi^2.
\end{equation}
Here $M$ is the mass of the black hole, $a$ its angular momentum per unit mass $(0\leq a \leq M)$, $\Sigma=r^2+a^2\cos^2 \theta$, and $\Delta=r^2-2Mr+a^2$. The event horizon is at $r=r_{+}$. Thorne \cite{thorne} has shown that $a\simeq 0.9984M$ is probably an upper limit to the rotation realisable by an accreting astrophysical black hole.

Scalar, electromagnetic and gravitational perturbations of this metric are all described by a master perturbation equation which, assuming a $t$ and $\phi$ dependence given by $e^{-i\omega t+im\phi}$, was separated by Teukolsky \cite{teukolsky}.
 
The separated differential equation for the angular part of the perturbation is 
\begin{equation}
(1-u^{2})\frac{d^{2}S_{s}}{du^{2}}-2u\frac{dS_{s}}{du}+W(u)S_{s}=0,
\end{equation}
where 
\begin{equation}
W=a^2\omega^2u^2-2a\omega su+\lambda +s-\frac{(m+su)^2}{1-u^2},
\end{equation}
and that for the radial part is
\begin{equation}
\Delta\frac{d^2R_{s}}{dr^2}+2(s+1)(r-M)\frac{dR_{s}}{dr}+V(r)R_{s}=0,
\end{equation}
where
\begin{equation}
V=\frac{K^2-2isK(r-M)}{\Delta}+4is\omega r-\lambda+2a\omega m-a^2\omega^2,
\end{equation}
with 
\begin{eqnarray}
u        & = & \cos\theta,\\
K        & = & (r^2+a^2)\omega-am.
\end{eqnarray}
The field spin-weight parameter $s$ takes the values $0£$, $-1$, $-2$, respectively, for outgoing scalar, electromagnetic, and gravitational waves. $\lambda$ is the angular separation constant for £$(2.2)$, and it reduces to $l(l+1)-s(s+1)$ at the Schwarzschild limit. 

For rotating black holes, we have to solve $(2.2)$ numerically following Leaver \cite{leaver}. Boundary conditions for $(2.2)$ are that $S_{s}$ is regular at the regular singular points $u=\pm 1$. The indices there are given by $\pm(m+s)/2$ at $u=1$ and $\pm(m-s)/2$ at $u=-1$. A solution to $(2.2)$ may be written as
\begin{equation}
S_{s}(u)=e^{a\omega u}(1+u)^{|m-s|/2}(1-u)^{|m+s|/2}\sum^{\infty}_{n=0}\hat a_{n}(1+u)^n.
\end{equation}
The series coefficients are related by a three term recurrence relation and the boundary condition at $u=1$ is satisfied only by its minimal solution sequence. The recurrence relation is
\begin{equation}
\hat\alpha_{0}\hat a_{1}+\hat\beta_{0}\hat a_{0}=0,
\end{equation}
\begin{equation}
\hat\alpha_{n}\hat a_{n+1}+\hat\beta_{n}\hat a_{n}
                     +\hat\gamma_{n}\hat a_{n-1}=0\;\;(n\geq 1),
\end{equation}
where $k_{1}=|m-s|/2$, $k_{2}=|m+s|/2$ and the coefficients of the recurrence relation are
\begin{eqnarray}
\hat\alpha_{n}&=& -2(n+1)(n+2k_{1}+1),\\
\hat\beta_{n} &=& n(n-1)+2n(k_{1}+k_{2}+1-2a\omega)-2a\omega(2k_{1}+s+1)\nonumber\\
              & & +(k_{1}+k_{2})(k_{1}+k_{2}+1)-(a^2\omega^2+s(s+1)+\lambda),\\
\hat\gamma_{n}&=&2a\omega(n+k_{1}+k_{2}+s).
\end{eqnarray}
The solution will be minimal if the angular separation constant $\lambda$ is a root of the continued fraction equation
\begin{equation}
0=\hat\beta_{0}-\frac{\hat\alpha_{0}\hat\gamma_{1}}{\hat\beta_{1}-}\frac{\hat\alpha_{1}\hat\gamma_{2}}{\hat\beta_{2}-}\frac{\hat\alpha_2\hat\gamma_3}{\hat\beta_{3}-}\frac{\hat\alpha_3\hat\gamma_4}{\hat\beta_{4}-}\cdots.
\end{equation}
A solution to the radial equation $(2.4)$ is found similarly to the solution for $(2.2)$. This equation has two regular singular points $r_{+}$ and $r_{-}$ as the roots of $\Delta$. We also define a new rotation parameter $b=(1-a^2/M^2)^{\frac{1}{2}}$, so that $b$ varies from $1$ to $0$ as $a$ varies from $0$ to $M$ (Kerr limit). Let us set $\hat a=a/2M$ and $\hat\omega=2M\omega$. The event horizon is at $r=r_{+}$. The indices at $r=r_{+}$ are $i\sigma_{+}$ and $-s-i\sigma_{+}$, where $\sigma_{+}=(\omega r_{+}-\hat a m)/b$. Since the second index corresponds to in-going radiation, we can establish on £$R_{s}$ the quasinormal boundary conditions
\begin{equation}
R_{s}\sim(r-r_{+})^{-s-i\sigma_{+}},\; as \;r\rightarrow r_{+},
\end{equation}
\begin{equation}
R_{s}\sim r^{-1-2s+i\omega}e^{i\omega r}, \; as \; r\rightarrow\infty.
\end{equation}
Thus, our solution can be expressed as
\begin{equation}
R_{s}=e^{i\omega r}(r-r_{-})^{-1-s+i\hat\omega+i\sigma_{+}}(r-r_{+})^{-s-i\sigma_{+}}\sum_{n=0}^{\infty}a_{n}\left(\frac{r-r_{+}}{r-r_{-}}\right)^n,
\end{equation}
where the expansion coefficients are again defined by a three term recurrence relation
\begin{equation}
\alpha_{0}a_{1}+\beta_{0}a_{0}=0,
\end{equation}
\begin{equation}
\alpha_{n}a_{n+1}+\beta_{n}a_{n}+\gamma_{n}a_{n-1}=0 \;\;(n\geq 1).
\end{equation}
The recursion coefficients are
\begin{eqnarray}
\alpha_{n}&=&n^2+(c_{0}+1)n+c_{0},\nonumber\\
\beta_{n} &=&-2n^2+(c_{1}+2)n+c_{3},\nonumber\\
\gamma_{n}&=&n^2+(c_{2}-3)n+c_{4}-c_{2}+2,
\end{eqnarray}
and the intermediate constants are given by
\begin{eqnarray}
c_{0}&=&1-s-i\hat\omega-\frac{2i}{b}\left(\frac{\hat\omega}{2}-\hat am\right),\nonumber\\
c_{1}&=&-4+2i\hat\omega(2+b)+\frac{4i}{b}\left(\frac{\hat\omega}{2}-\hat am\right),\nonumber\\
c_{2}&=&s+3-3i\hat\omega-\frac{2i}{b}\left(\frac{\hat\omega}{2}-\hat am\right),\nonumber\\
c_{3}&=&\hat\omega^2(4+2b-\hat a^2)-2\hat am\hat\omega-s-1+(2+b)i\hat\omega-\lambda+\frac{4\hat\omega+2i}{b}\left(\frac{\hat\omega}{2}-\hat am\right),\nonumber\\
c_{4}&=&s+1-2\hat\omega^2-(2s+3)i\hat\omega-\frac{4\hat\omega+2i}{b}\left(\frac{\hat\omega}{2}-\hat am\right),
\end{eqnarray}
The radial series solution converges and the $r=\infty$ boundary condition is satisfied if, for a given $a$, $m$, $\lambda$, and $s$, the frequency $\omega$ is a root of the continued fraction equation
\begin{equation}
0=\beta_{0}-\frac{\alpha_{0}\gamma_{1}}{\beta_{1}-}\frac{\alpha_{1}\gamma_{2}}{\beta_{2}-}\frac{\alpha_{2}\gamma_{3}}{\beta_{3}-}\cdots,
\end{equation}
or any of its inversions.
 
It is useful to bear in mind the behaviour of the quasinormal frequencies of Kerr black holes under complex conjugation. Looking at equations $(2.2)$ and $(2.4)$ we can verify that, if $\rho_{n,m}=-i\omega_{n,m}$ and $\lambda_{l,m}$ are a quasinormal frequency and the corresponding angular eigenvalue, then $\rho_{n,-m}=\rho^{\ast}_{n,m}$ and $\lambda_{l,-m}=\lambda^{\ast}_{l,m}$. This satisfies the requirement of complex conjugate quasinormal modes, because the perturbative equation was separated by Teukolsky \cite{teukolsky}, writing the wave equation as
\begin{equation}
\Psi(t,r,\theta,\phi)=\frac{1}{2\pi}\int e^{-iwt}\sum_{l=|s|}^{\infty}\sum_{m=-l}^{l}e^{im\phi} S_{s}(u)R_{s}(r)d\omega,
\end{equation}
and in this equation the sum is over both positive and negative values of $m$.
\section{New technique for the angular equation}
The angular separation constant can be approximated for small $a\omega$ by perturbation theory. This is, for example, the method used by Sasaki in \cite{sasaki}. The angular eigenvalues are, in that case, expanded in powers of $a\omega$,
\begin{equation}
\lambda=\lambda_0+a\omega\lambda_1+a^2\omega^2\lambda_2+O((a\omega)^3).
\end{equation}
Explicit values of the coefficients are
\begin{eqnarray}
\lambda_0&=&l(l+1)-2=(l-1)(l+2),\nonumber \\
\lambda_1&=&-2m\frac{l(l+1)+4}{l(l+1)},\nonumber \\
\lambda_2&=&-2(l+1)(c^{l+1}_{lm})^2+2l(c^{l-1}_{lm})^2+\frac{2}{3}-\frac{2}{3}\frac{(l+4)(l-3)(l^2+l-3m^2)}{l(l+1)(2l+3)(2l-1)},\nonumber
\end{eqnarray}
with
\begin{eqnarray}
c^{l+1}_{lm}&=&\frac{2}{(l+1)^2}\left[\frac{(l+3)(l-1)(l+m+1)(l-m+1)}{(2l+1)(2l+3)}\right]^{1/2},\nonumber\\
c^{l-1}_{lm}&=&-\frac{2}{l^2}\left[\frac{(l+2)(l-2)(l+m)(l-m)}{(2l+1)(2l-1)}\right]^{1/2}. \nonumber
\end{eqnarray}
This method works as far as the expansion parameter $a\omega$ is sufficiently small, so that if removes the possibility of dealing with high frequencies. Since we are interested both in slowly and rapidly damped modes, we prefer not to have such a limitation. 
As already mentioned in the introduction, we follow Leaver's continued fraction method and we adopt a new strategy for the numerical solution of the two continued fraction equations $(2.14)$ and $(2.22)$. Instead of a two dimensional root finding routine, which gives simultaneously frequencies and separation constants in the complex plane, we first compute the angular eigenvalues and then use those values in the continued fraction coming out from the radial solution.

Firstly, we write a computer program able to approximate the continued fraction $(2.14)$ in the most efficient way and calculate the roots, which are the angular eigenvalues. We use an excellent method known as {\em modified Lent's algorithm} \cite{flannery}, which essentially converts the continued fraction to a quotient of two series
\begin{equation}
f_{n}=\frac{A_{n}}{B_{n}},
\end{equation}
where $A_{n}$ and $B_{n}$ are determined from recurrence relations connected to the the coefficients of the continued fraction. The series are evaluated from left to right, and the algorithm is stopped when the value of the absolute value of the difference between two consecutive fractions is sufficiently small.
Within the same program, the roots are searched by the use of the complex one dimensional Newton-Rapson method routine.

Then, we fit these numerical data for $\lambda$ to a Pade' approximation for {\em real $a\omega$} given by
\begin{equation}
\lambda=\frac{a_{0}+a_{1}a\omega+a_{2}(a\omega)^2+a_{3}(a\omega)^3+a_{4}(a\omega)^4}{1+b_{1}a\omega+b_{2}(a\omega)^2+b_{3}(a\omega)^3}.
\end{equation}
Finally, the Pade' approximation and numerical data are compared for {\em imaginary $a\omega$}.

In the Pade' formula $(3.3)$, the numerator is one degree higher than the denominator. The reason for this choise is that, the Teukolsky's angular equation belongs to the family of differential equations for angular prolate spheroidal wave functions. The behaviour of the angular eigenvalues, for large $a\omega$, related to this kind of equation is well known \cite{abramowitz}.

Fitting simulations have been done for different values of the angular parameter $a$, for several values of the spherical harmonic parameters $l$ and $m$ and they always work accurately when we compare our results with Leaver's.
In Fig.1, 
\begin{figure}
\begin{center}
\begin{tabular}{cc}
  \psfrag{Re(lambda)}{$\lambda_R$}
  \psfrag{omega}{$\omega$}
  \epsfysize=6.0cm
  \epsfbox{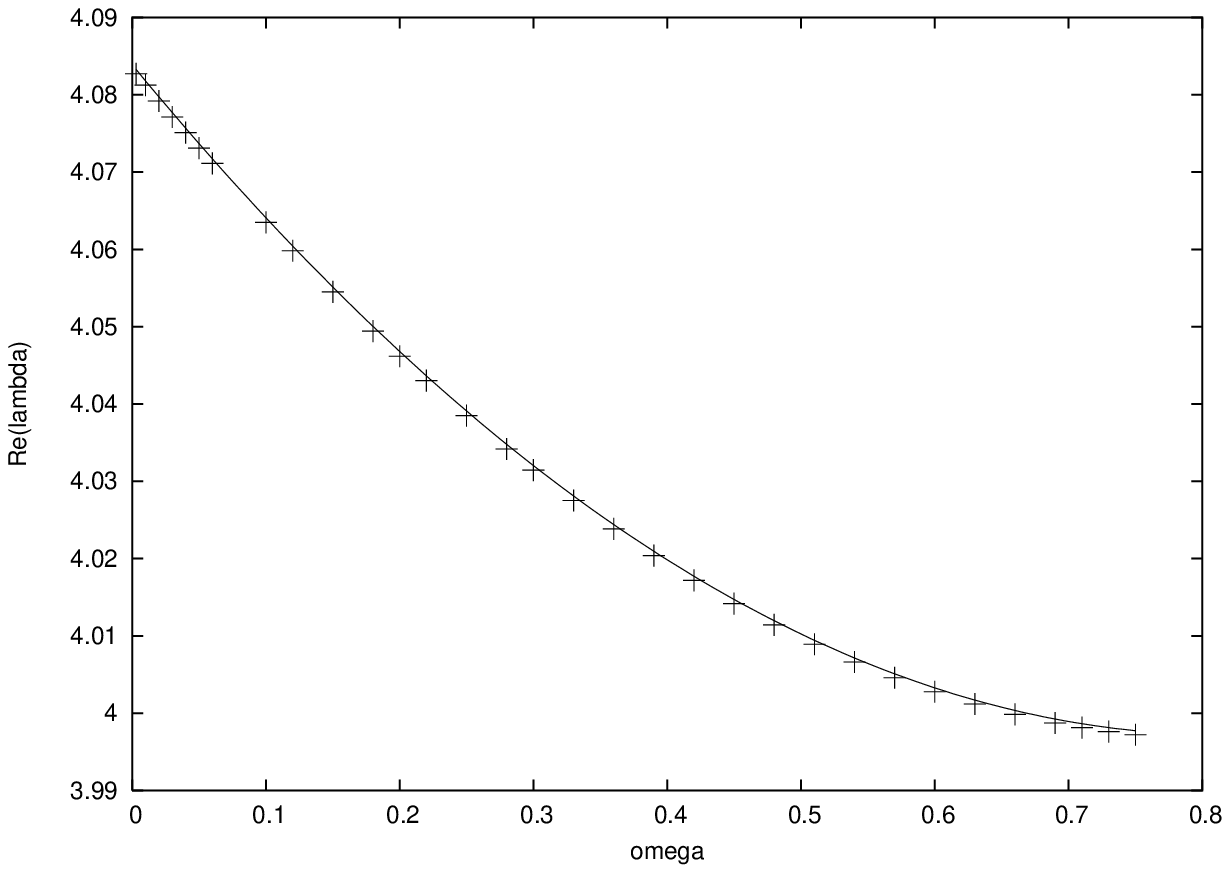} &
  \psfrag{Im(lambda)}{$\lambda_I$}
  \psfrag{omega}{$\omega$}
  \epsfysize=6.0cm
  \epsfbox{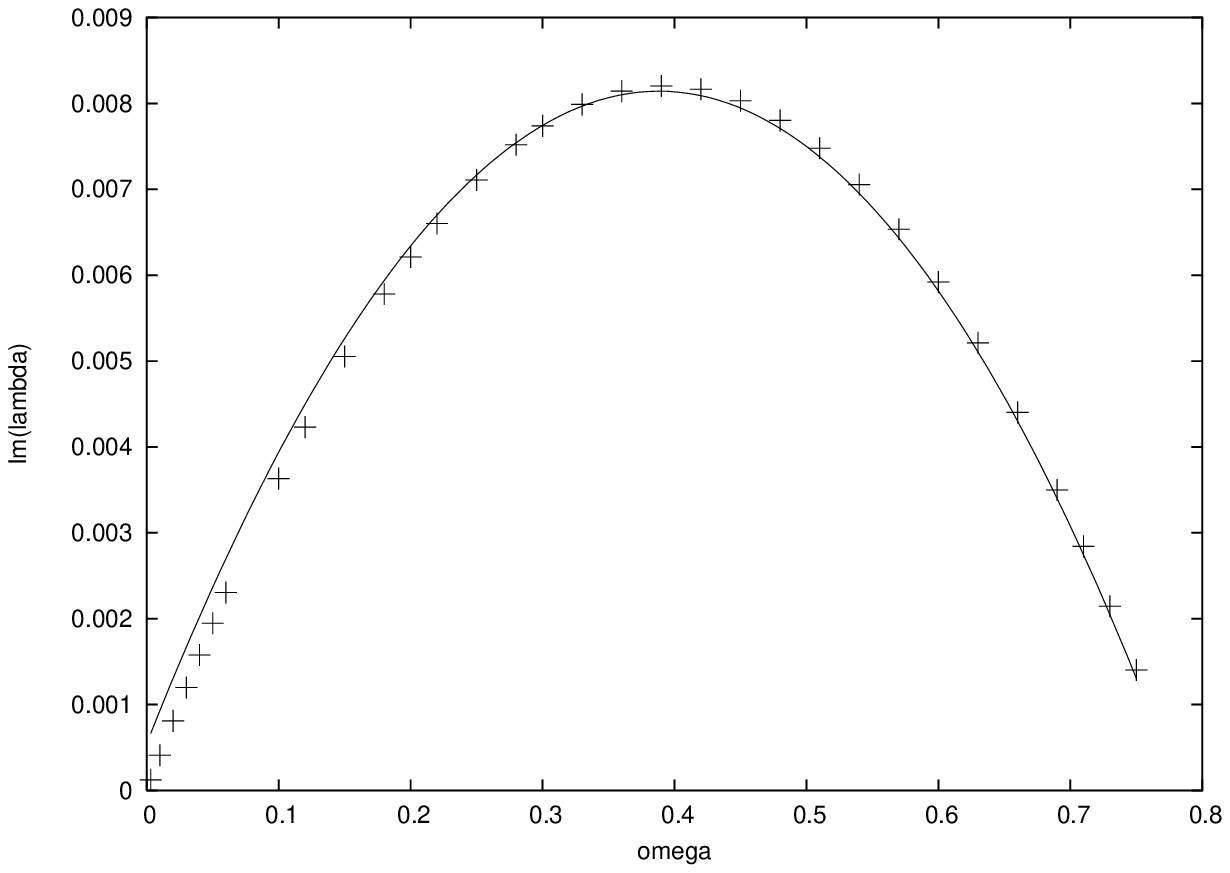} \\
  \psfrag{Re(lambda)}{$\lambda_R$}
  \psfrag{omega}{$\omega$}
  \epsfysize=6.0cm
  \epsfbox{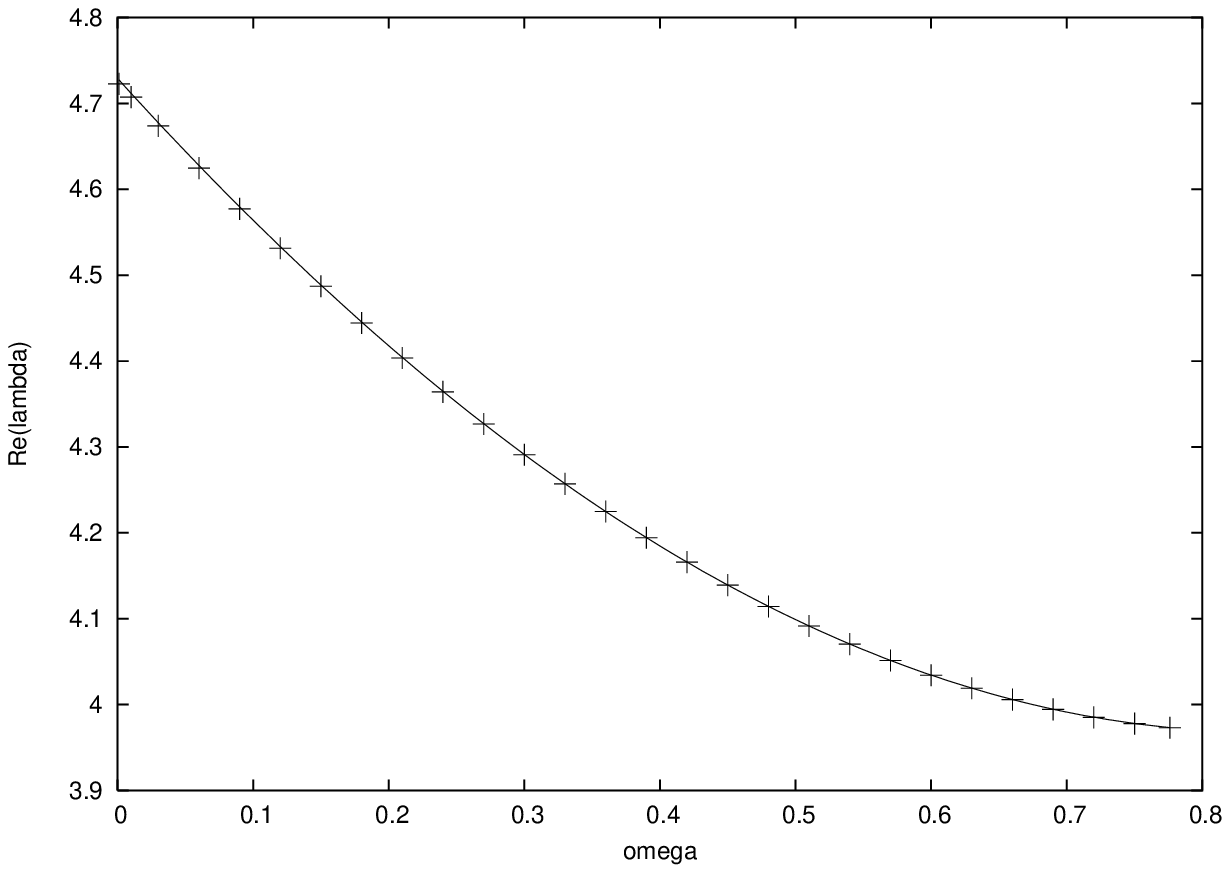} &
  \psfrag{Im(lambda)}{$\lambda_I$}
  \psfrag{omega}{$\omega$}
  \epsfysize=6.0cm
  \epsfbox{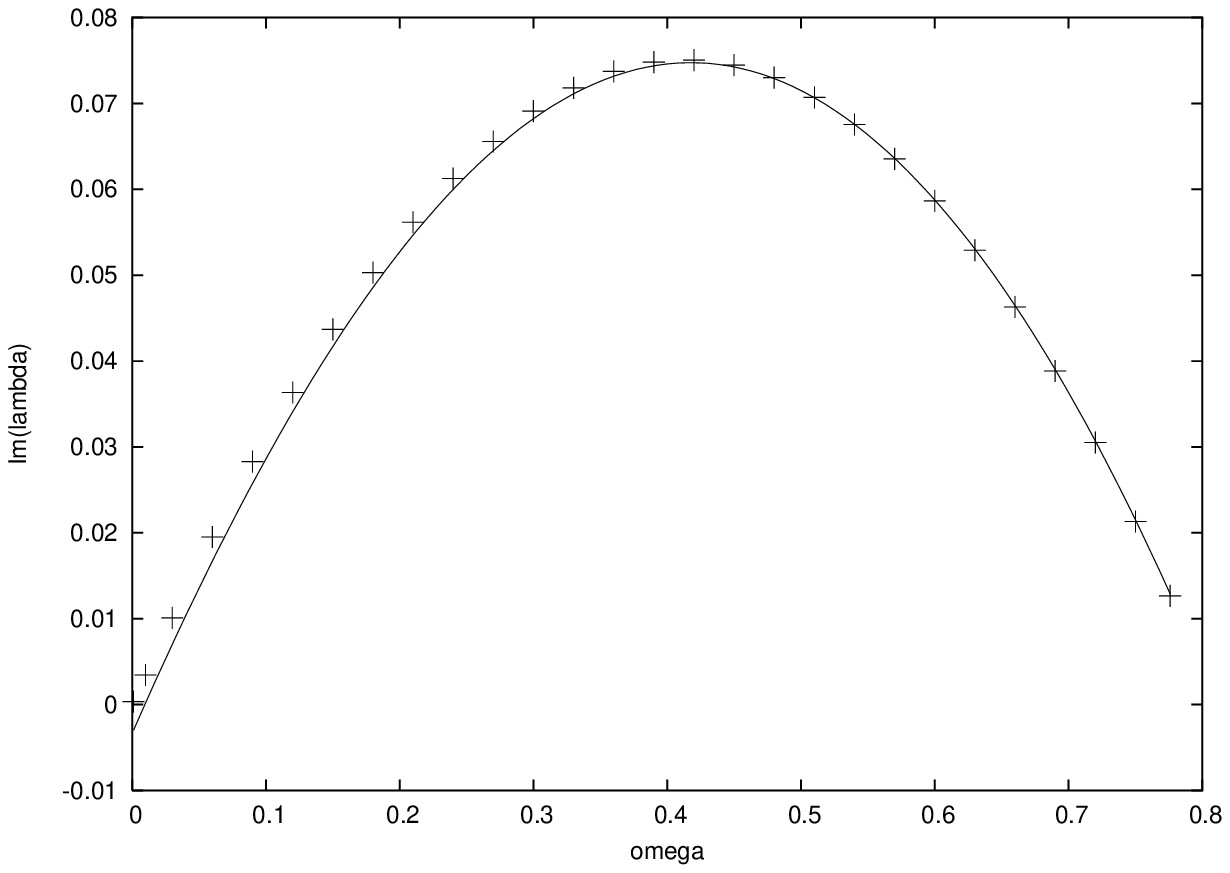} \\
  \psfrag{Re(lambda)}{$\lambda_R$}
  \psfrag{omega}{$\omega$}
  \epsfysize=6.0cm
  \epsfbox{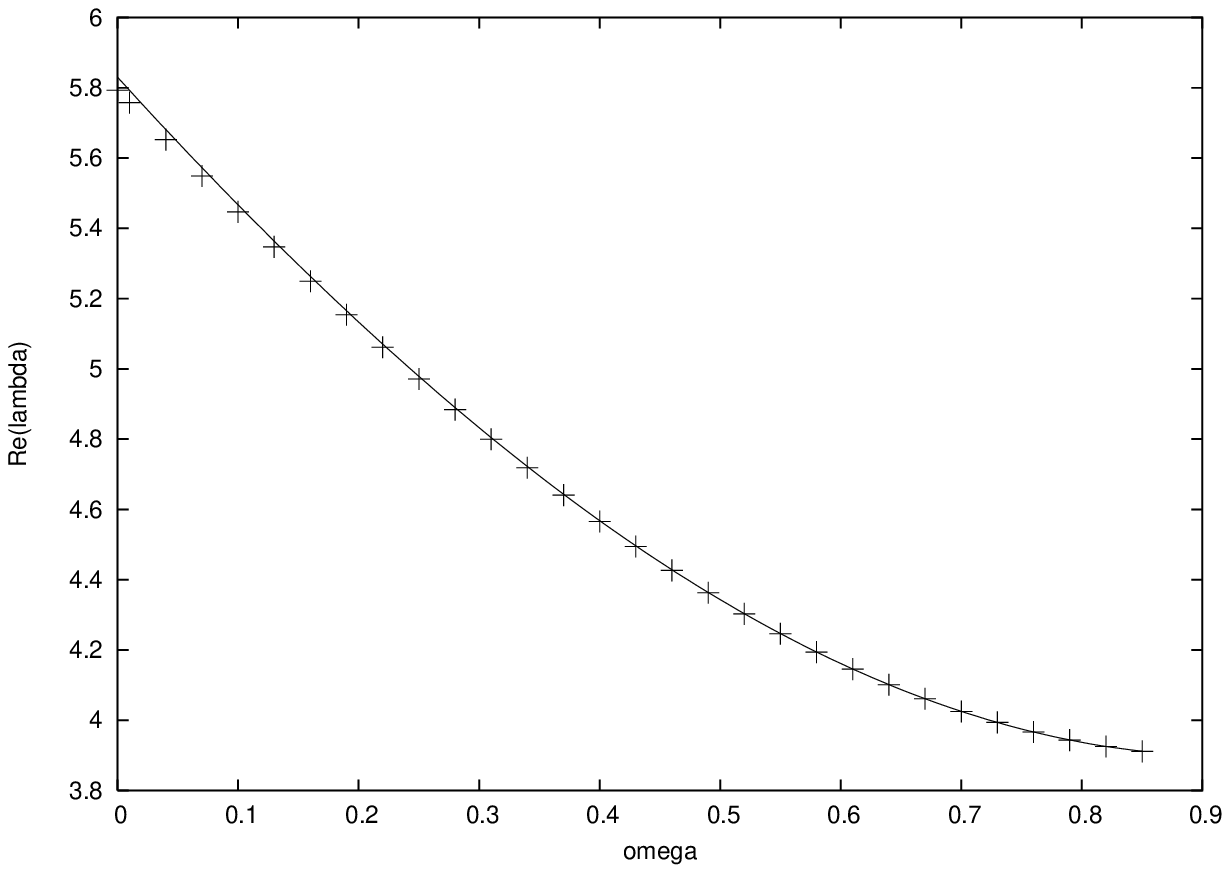} &
  \psfrag{Im(lambda)}{$\lambda_I$}
  \psfrag{omega}{$\omega$}
  \epsfysize=6.0cm
  \epsfbox{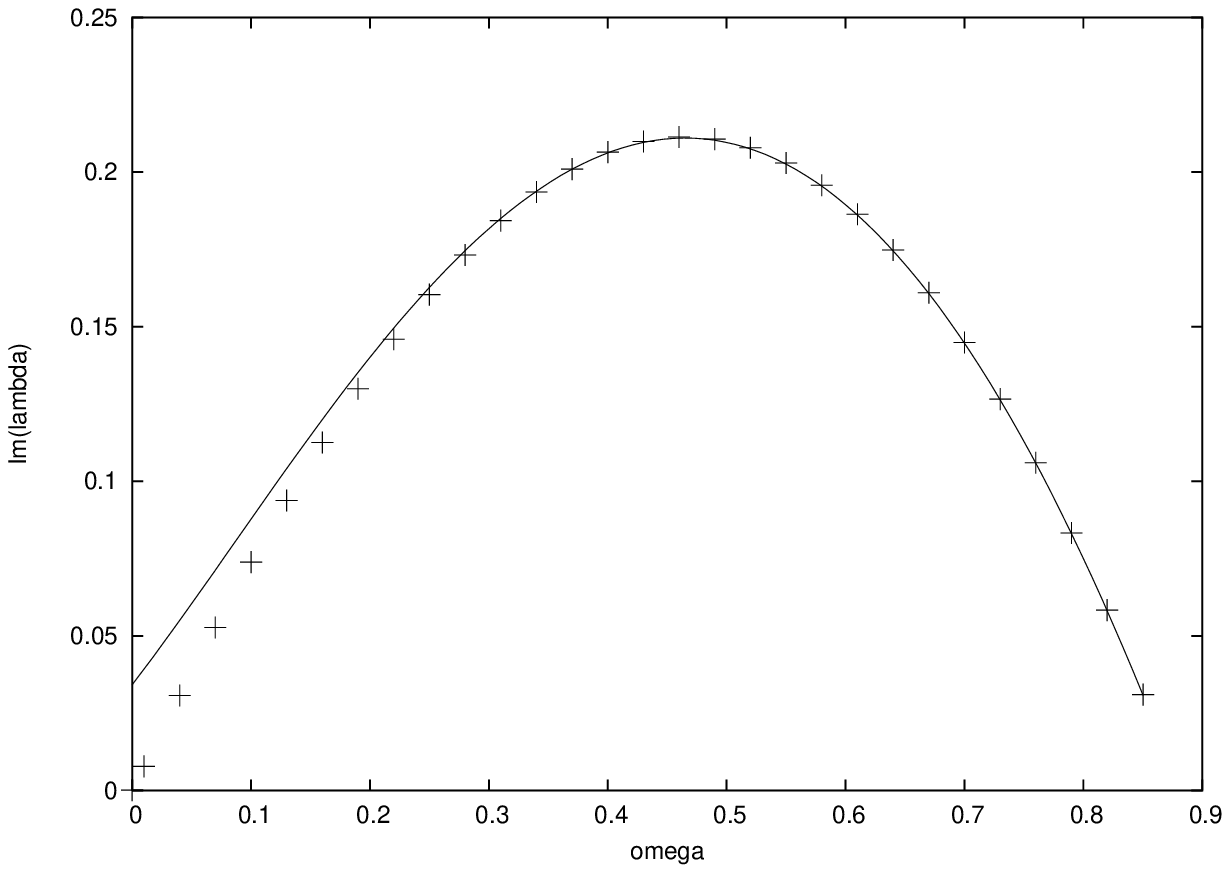} \\
\end{tabular}
\caption{The real and imaginary parts of the Pade' approximation of the angular eigenvalue $\lambda$ are shown on the left and right columns respectively. They are plotted as functions of $\omega$. These figures represent results we get for $l=2$, $m=0$ and with the angular parameter $\hat a=0.1$, $0.3$, $0.49$ from the top couple of plots to the bottom one. Note that, the Pade' fit of the numerical data for $\lambda$ has been done for real values of $\omega$.}
\end{center}
\end{figure}
we show how well our Pade' approximation works for $m=0$ and $l=2$ both for the real and imaginary parts of $\lambda$.
The fitting routine we have used is {\em gnufit}, but the same results can be obtained using nonlinear curve fitting packages in {\em Maple}, available on the web. The fit parameters of the rational function representing $\lambda$ are shown in Table $1$ for $m=0$ and in Table $2$ and Table $3$ for $m=\pm 1$.

\begin{table}[h!]
\begin{tabular}{l|cccccccc}
\hline
\hline
$\hat a$  &$a_{0}$&$a_{1}$    & $a_2$   &$a_3$    &$a_4$     &$b_1$    &$b_2$    &$b_3$  \\ \hline
0.0&4.0&0.0&0.0&0.0&0.0&0.0&0.0&0.0\\
0.1&4.0&-0.2979 &-0.3994&0.0420&-0.0294 &-0.0744 &0.0310&0.0009 \\ 
0.2&4.0&-0.3127&-0.3939&0.0434 &-0.03   &-0.0781&0.0323 &0.0008\\
0.3&4.0&-0.5886&-0.2857&0.0679 &-0.0396 &-0.1470&0.0588 &-0.0009\\
0.4&4.0&-0.5284&-0.2889&0.0619 &-0.04038&-0.1320&0.0582 &-0.0008\\
0.49&4.0&-0.5378&-0.2872&0.0627&-0.0404&-0.1343&0.0586&-0.00086\\
\hline
\hline
\end{tabular}
\caption{This table shows, for $m=0$, the parameters we obtain by fitting the Pade' approximation for $\lambda$, to the numerical data coming out from the computer program for the calculation of the angular eigenvalues of Teukolsky equation.}
\end{table}
\begin{table}[!ht]
\begin{tabular}{l|cccccc}
\hline
\hline
$\hat a$  &$a_{0}$&$a_{1}$    & $a_2$   &$a_3$     &$b_1$    &$b_2$ \\ \hline
0.0&4.0&0.0&0.0&0.0&0.0&0.0\\
0.1&4.0&0.8550 &-0.8523&0.0792&-0.1195&-0.0289 \\ 
0.2&4.0&2.2571&-0.4582&-0.0919&0.2284&-0.0380\\
0.3&4.0&2.8878&-0.3198&-0.1651&0.3838&-0.0483\\
0.4&4.0&3.2568&-0.2402&-0.2078&0.4747&-0.0546\\
0.49&4.0&3.2944&-0.2335&-0.2118&0.4838&-0.0553\\
\hline
\hline
\end{tabular}
\caption{This table shows, for $m=-1$, the parameters we obtain by fitting the Pade' approximation for $\lambda$, to the numerical data coming out from the computer program for the calculation of the angular eigenvalues of Teukolsky equation.}
\end{table}
\begin{table}[ht!]
\begin{tabular}{l|cccccc}
\hline
\hline
$\hat a$  &$a_{0}$&$a_{1}$    & $a_2$   &$a_3$     &$b_1$    &$b_2$ \\ \hline
0.0&4.0&0.0&0.0&0.0&0.0&0.0\\
0.1&4.0&-4.1751&0.5889&0.2868&-0.7104&0.0543\\
0.2&4.0&-2.1949&-0.2201&0.0469&-0.2151&0.0163\\
0.3&4.0&-2.7029&-0.0173&0.1112&-0.3423&0.0252\\
0.4&4.0&-2.5610&-0.0748&0.0937&-0.3068&0.0226\\
0.49&4.0&-2.7425&-0.0014&0.1161&-0.3522&0.0258\\
\hline
\hline
\end{tabular}
\caption{This table shows, for $m=+1$, the parameters we obtain by fitting the Pade' approximation for $\lambda$, to the numerical data coming out from the computer program for the calculation of the angular eigenvalues of Teukolsky equation.}
\end{table}
\section{Radial equation and numerical results}
As a consequence of the new technique, the angular equation with its angular eigenvalues is not an issue anymore. We are in the position of carrying on our numerical search for quasinormal frequencies and we switch all our attention at the radial equation.

As already done in the angular case, we must efficiently represent the radial continued fraction $(2.22)$. It is convenient in this case if we consider
 the $n$th inversion of the radial continued fraction, that is 
\begin{equation}
0=\left[\hat\beta_{n}-\frac{\hat\alpha_{n-1}\hat\gamma_{n}}{\hat\beta_{n-1}-}\frac{\hat\alpha_{n-2}\hat\gamma_{n-1}}{\hat\beta_{n-2}-}\cdots \frac{\hat\alpha_{0}\hat\gamma_{1}}{\hat\beta_{0}}\right]-\left[ \frac{\hat\alpha_{n}\hat\gamma_{n+1}}{\hat\beta_{n+1}-}\frac{\hat\alpha_{n+1}\hat\gamma_{n+2}}{\hat\beta_{n+2}-}\cdots \right],
\end{equation}
because the $n$th quasinormal frequency is usually found to be numerically the most stable root of the $n$th inversion \cite{leaver}.
We, then, use again {\em Lentz's method} both for the finite and the infinite fractions in $(4.1)$. Finally, we write another simple computer program similar to the one used for the angular equation, able to approximate the sum $(4.1)$ and at the same time to compute its roots. In this second program we make use of the Pade' rational function to represent the angular eigenvalues which appear in $(2.22)$. In this case the roots are obviously the Kerr quasinormal frequencies we are looking for.

Using this new approach we actually recover former results up to the ninth mode\cite{leaver},\cite{onozawa} and in some cases we are able to go even beyond them.
In Fig. 2, 
\begin{figure}
\begin{center}
\begin{tabular}{cc}
  \psfrag{a = 0.0}{$\hat a = 0.0$}
  \psfrag{Im(w)}{$\omega_I$}
  \epsfysize=6.0cm
  \epsfbox{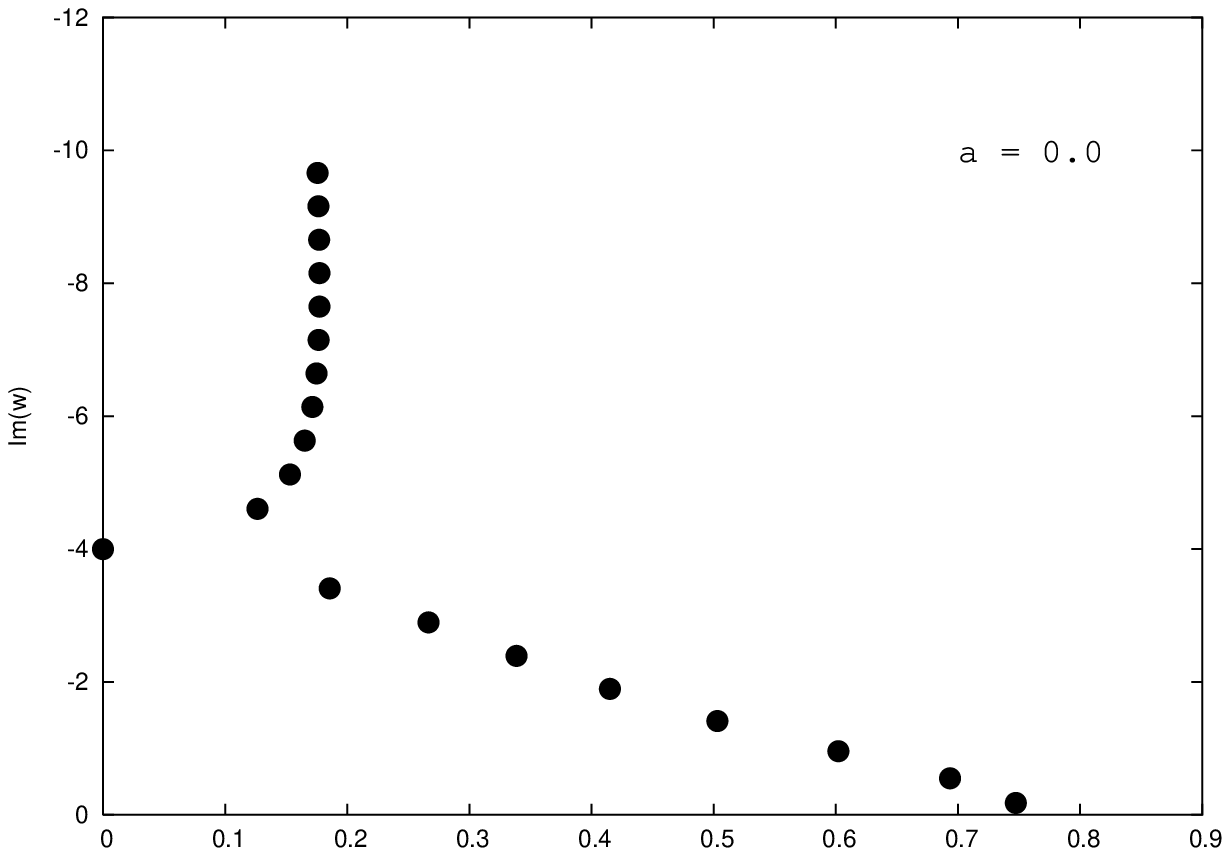} &
  \psfrag{a = 0.1}{$\hat a = 0.1$}
  \epsfysize=6.0cm
  \epsfbox{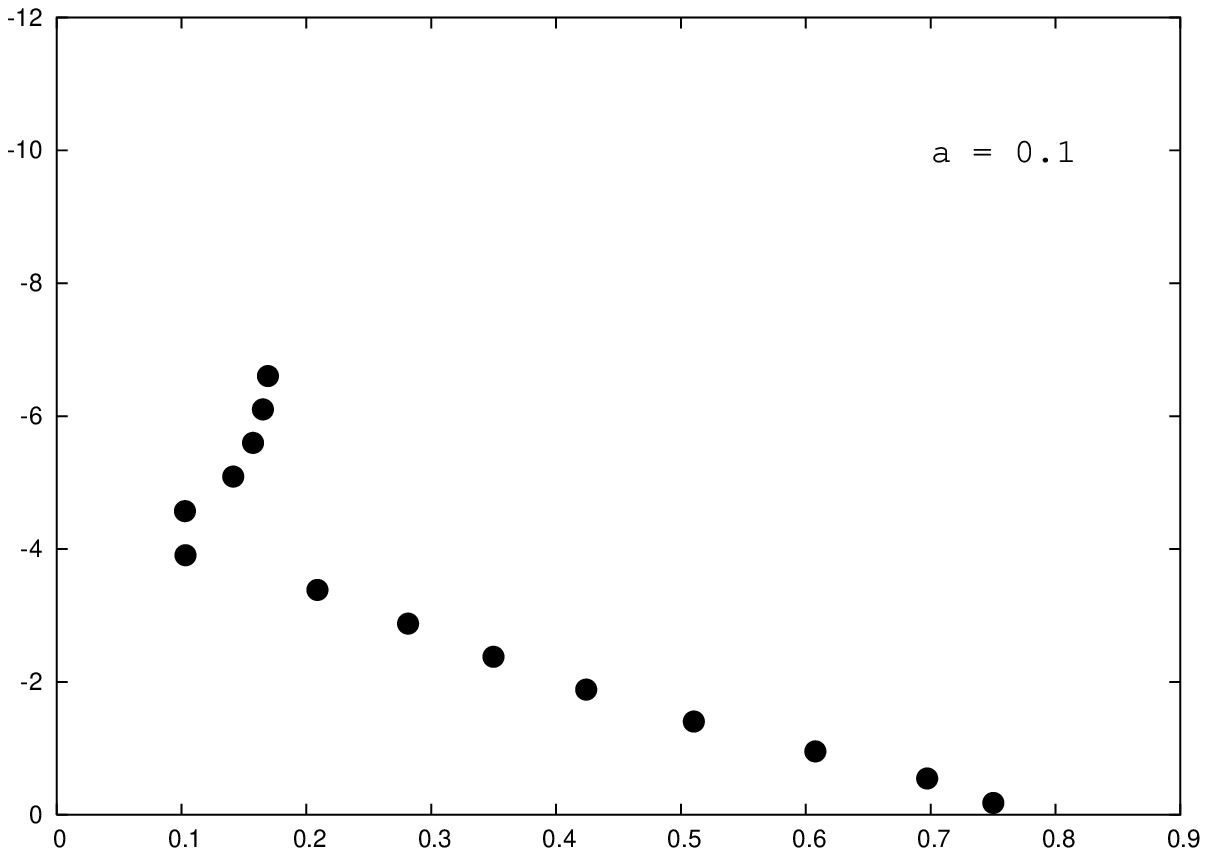} \\
  \psfrag{a = 0.2}{$\hat a = 0.2$}
  \psfrag{Im(w)}{$\omega_I$}
  \epsfysize=6.0cm
  \epsfbox{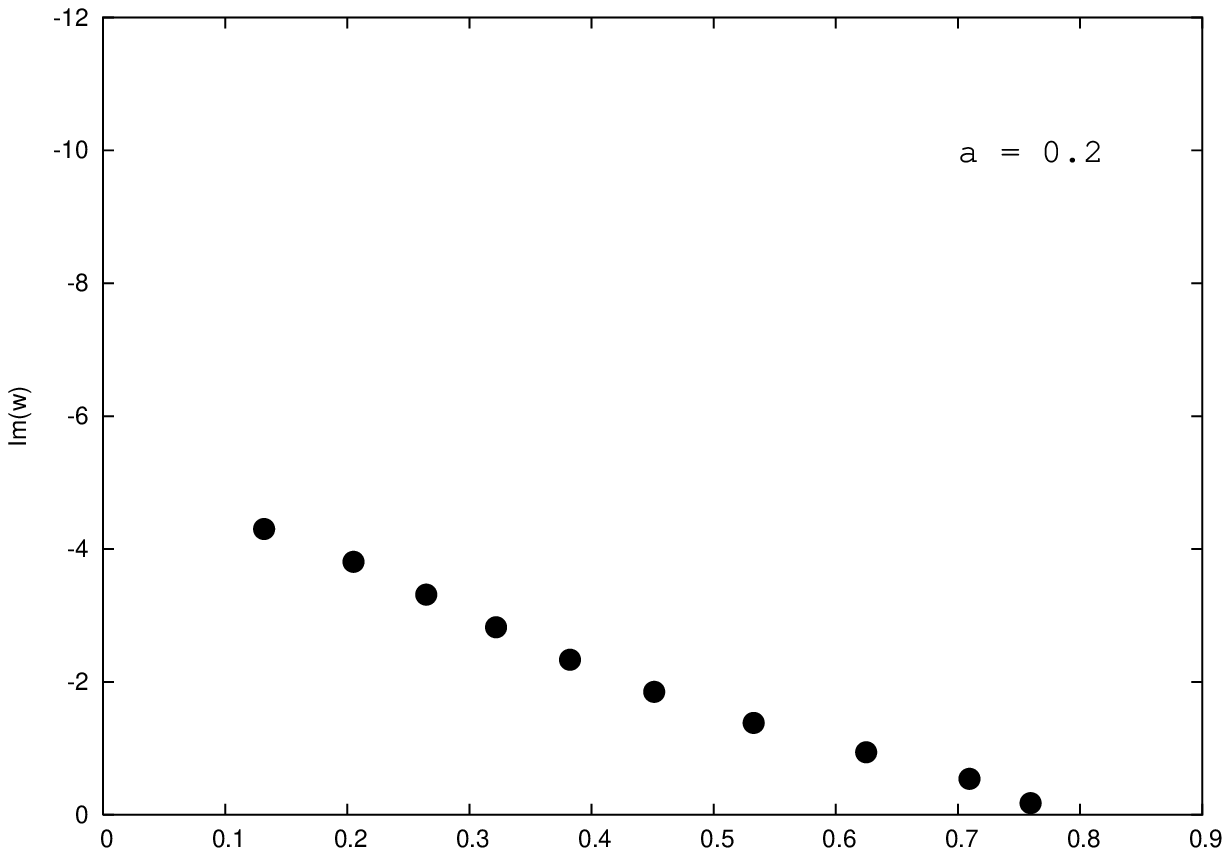} &
  \psfrag{a = 0.3}{$\hat a = 0.3$}
  \epsfysize=6.0cm
  \epsfbox{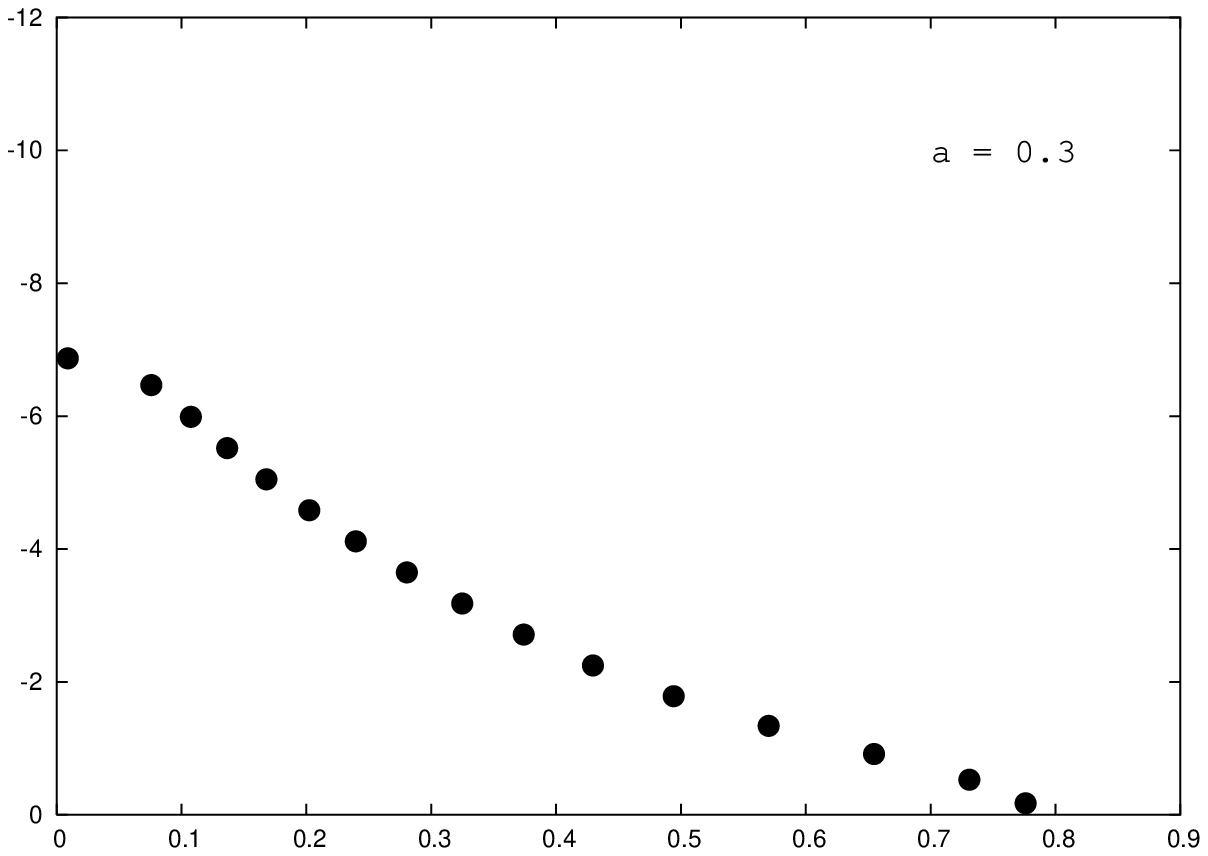} \\
  \psfrag{a = 0.4}{$\hat a = 0.4$}
  \psfrag{Re(w)}{$\omega_R$}
  \psfrag{Im(w)}{$\omega_I$}
  \epsfysize=6.0cm
  \epsfbox{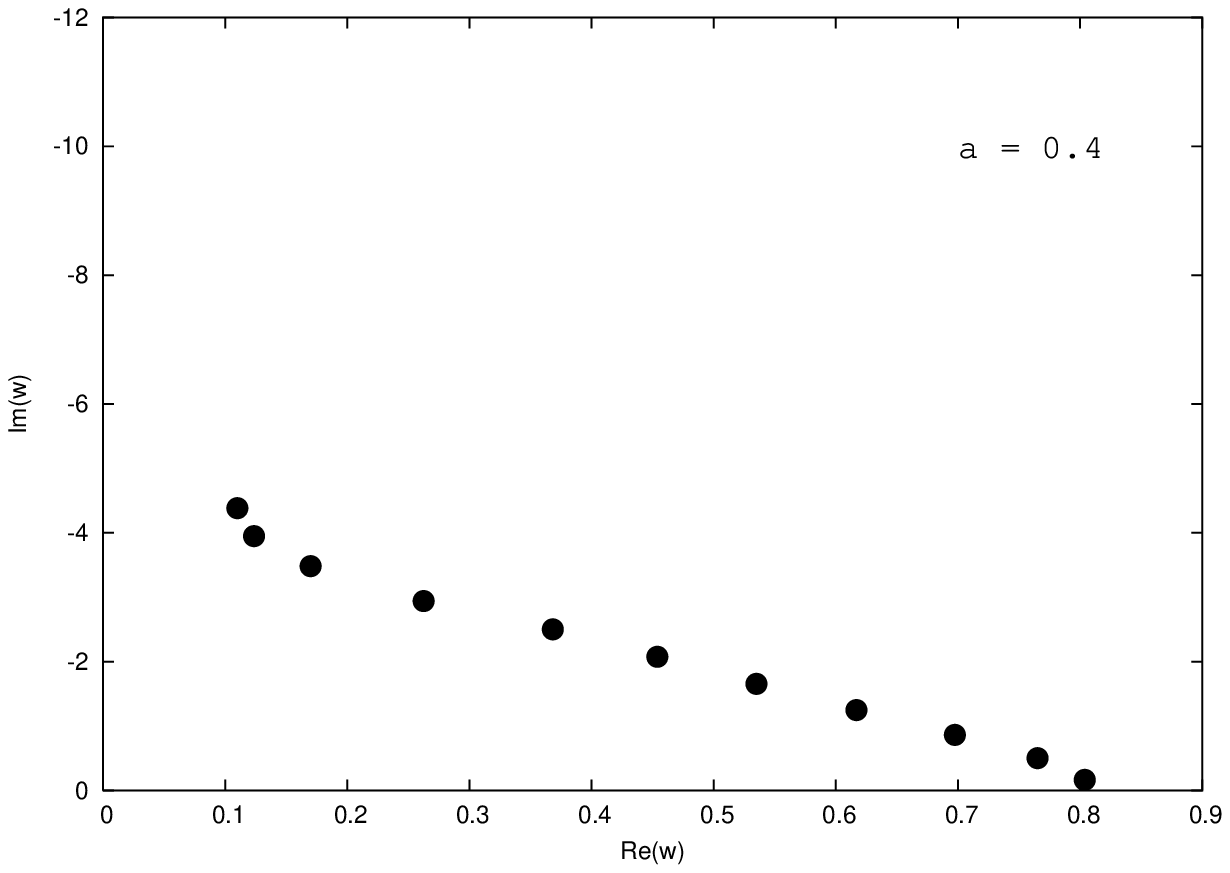} &
  \psfrag{a = 0.49}{$\hat a = 0.49$}
  \psfrag{Re(w)}{$\omega_R$}
  \epsfysize=6.0cm
  \epsfbox{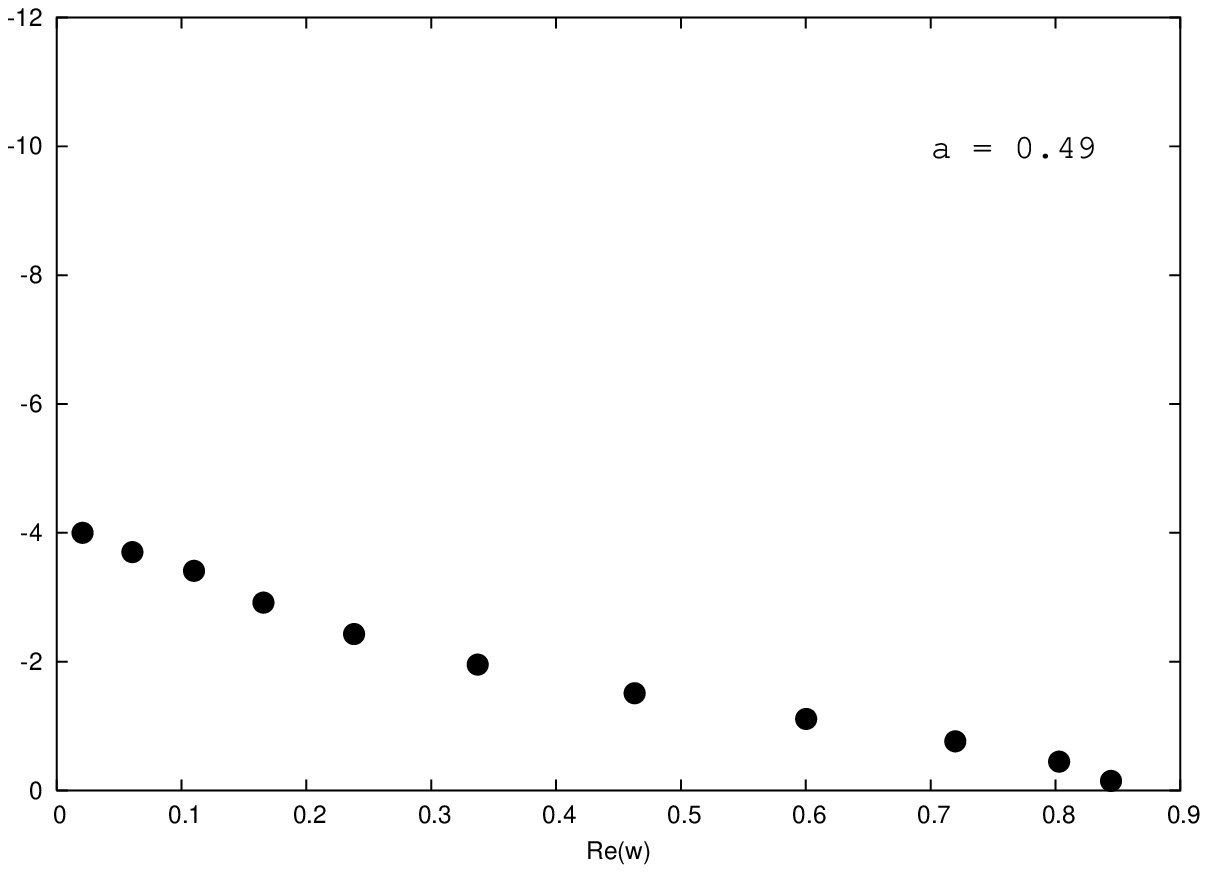} \\
\end{tabular}
\caption{First quasinormal frequencies for $l=2$ and $m=0$ scaled by $2M$, represented in the complex plane. In each box we have a different value for the angular momentum per unit mass $a$, which ranges between $0.0$ and $M$ (very close to the Kerr limit).}
\end{center}
\end{figure}
we have some plots showing patterns of quasinormal frequencies for the angular parameter $\hat a$ ranging from $\hat a=0.1$ to $\hat a=0.49$, which is very close to the Kerr limit. The top left box represents the frequencies for $\hat a=0.0$, $l=2$, and $m=0$ and it is the classical result for Schwrzschild black holes. It can be found, for instance, in \cite{nollert},\cite{kokkotas}.
In Fig. 3, 
\begin{figure}
\begin{center}
 \leavevmode
 \psfrag{Re(w)}{$\omega_R$}
 \psfrag{Im(w)}{$\omega_I$}
 \epsfysize=11.0cm
 \epsfbox{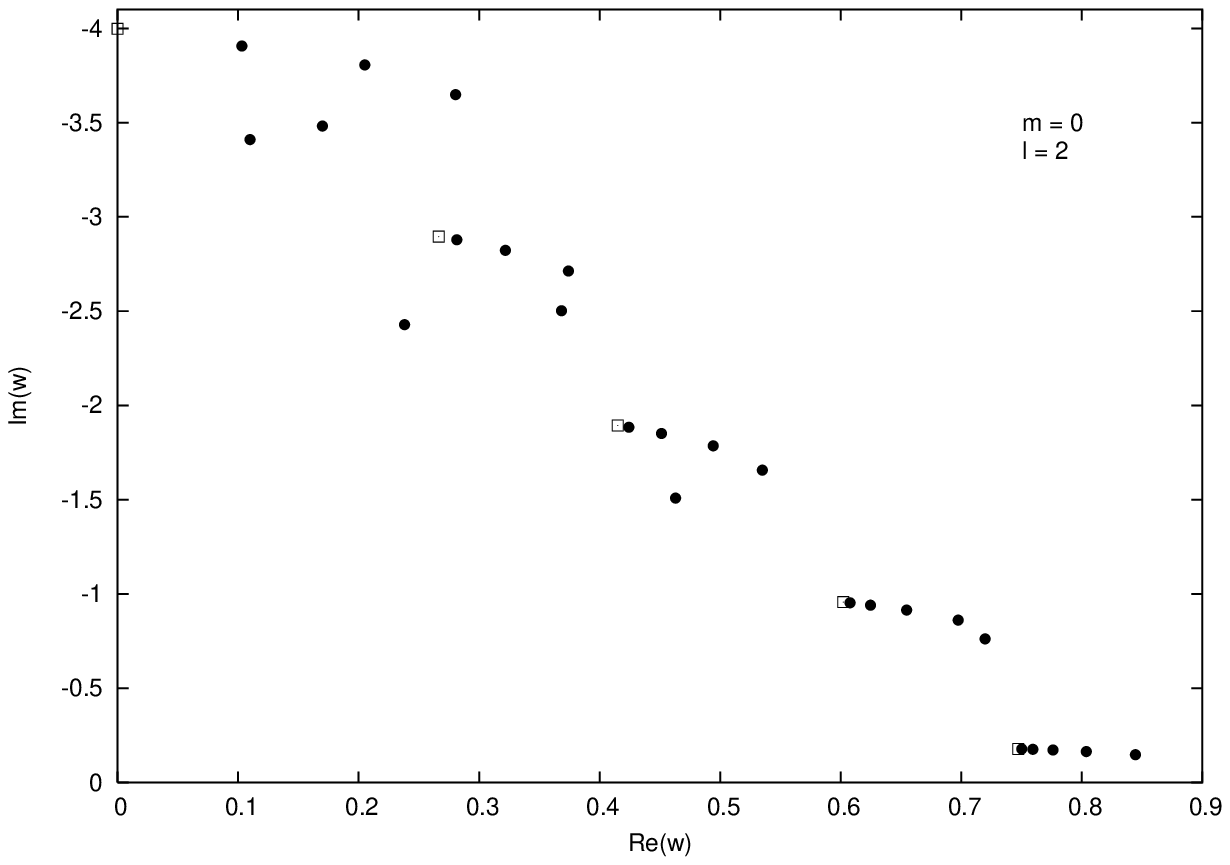}
 \psfrag{Re(w)}{$\omega_{R}$}
 \psfrag{Im(w)}[90]{$\omega_{I}$}
\caption{The first, third, fifth, seventh and ninth quasinormal modes for $l=2$ and $m=0$, scaled by $2M$, are shown. The single modes are paramtrised by the value of the rotation parameter $\hat a$. The left upper end of each mode (squares), represents the frequency at the Schwarzschild limit $\hat a=0$. The right end in the first and third modes and the lower left end in the others, represent the frequency at $\hat a=0.49$. The intermediate dots, correspond to $\hat a=0.1, 0.2, 0.3, 0.4$. As clearly shown, the ninth mode is on the imaginary axis for $\hat a=0.0$, after that it moves away as expected.}
\end{center}
\end{figure}
we show details of the modes on a different scale. Parametrisation is done by the the value of the rotation parameter $a$. It looks clear that the ninth mode is initially on the imaginary axis as expected.

In Fig. 4,
\begin{figure}
\begin{center}
 \leavevmode
 \psfrag{Re(w)}{$\omega_R$}
 \psfrag{Im(w)}{$\omega_I$}
 \epsfysize=11.0cm
 \epsfbox{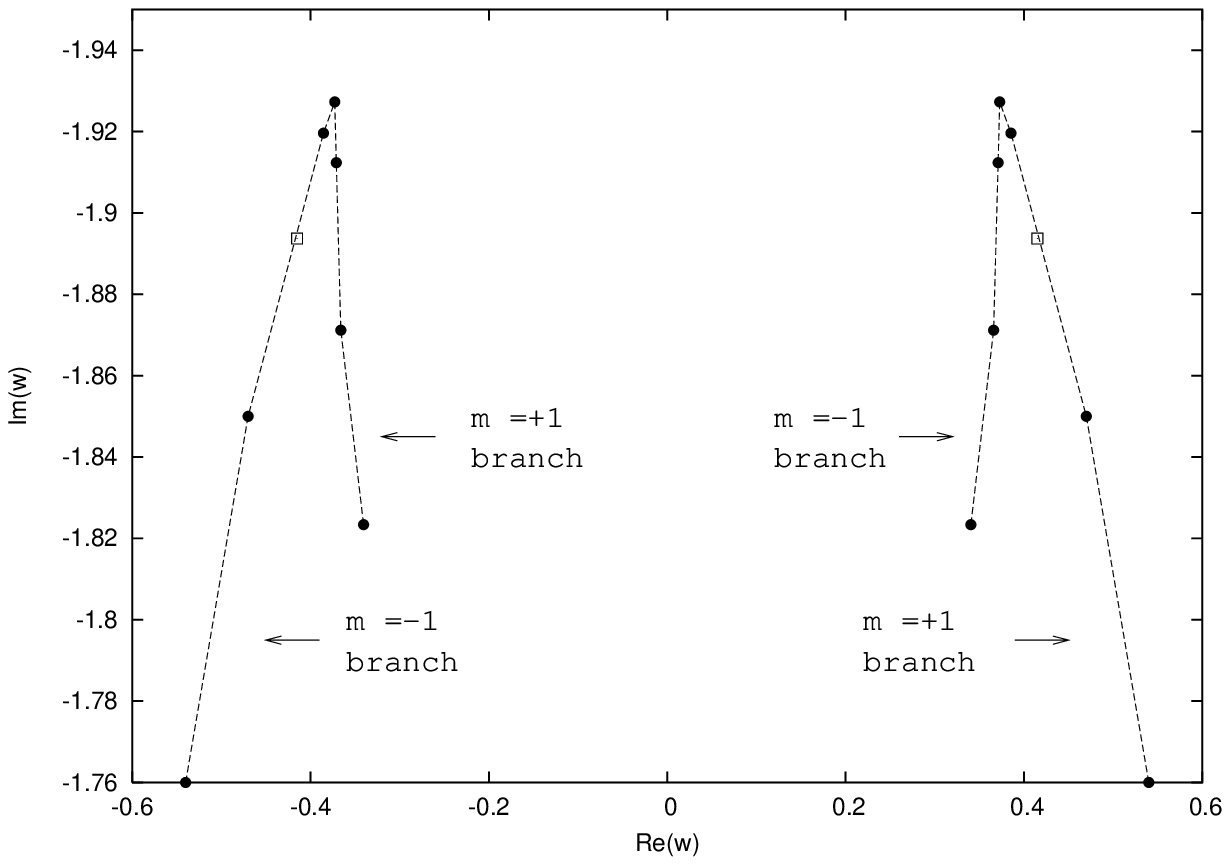}
\caption{A detail of the curve representing the fifth mode for $m=\pm 1$ is plotted. We also illustrate the complex conjugate nature of the Kerr quasinormal frequencies. The squares represent the complex conjugate frequencies for $a=0.0$.}
\end{center}
\end{figure}
we plot a detail of the curve representing the fifth mode for $m=\pm 1$ and we illustrate the complex conjugate symmetry of the Kerr quasinormal frequencies mentioned at the end of section $2$. This figure, shows again, that our technique reproduces Leaver's results accurately. The peak we get for $m=-1$ and angular parameter $\hat a$ ranging between zero and the Kerr limit, confirms similar behaviour found in Leaver \cite{leaver}.

In Table $3$, 
\begin{table}
\begin{tabular}{l|ccccc}
\hline
\hline
$\hat a$&$\hat\omega_1$&$\hat\omega_2$&$\hat\omega_3$&$\hat\omega_4$&$\hat\omega_5$\\ \hline
0.0&(0.747343, -0.177925)&(0.693422, -0.547829)&(0.602107, -0.956555)&(0.503011, -1.4103)&(0.415028, -1.89369)\\
0.1&(0.750252, -0.177401)&(0.697296, -0.546029)&(0.607671, -0.952749)&(0.510396, -1.40365)&(0.424195, -1.88378)\\
0.2&(0.75936, -0.175653)&(0.709343, -0.540027)&(0.624762, -0.940155)&(0.532732, -1.38181)&(0.451287, -1.85134)\\
0.3&(0.776104, -0.17199)&(0.731061, -0.527509)&(0.654649, -0.914241)&(0.57023, -1.33741)&(0.494169, -1.78589)\\
0.4&(0.803834, -0.164314)&(0.765136, -0.501476)&(0.6975, -0.86143)&(0.616905, -1.24857)&(0.535028, -1.65631)\\
0.49&(0.844509, -0.147065)&(0.802873, -0.446104)&(0.7198, -0.7616)&(0.60027, -1.11018)&(0.462988, -1.50839)\\
\hline
\hline
\end{tabular}
\caption{In this table, we show the numerical values of the first five Kerr quasinormal frequencies for $l=2$, $m=0$ and $\hat a=0.0,..,0.49$.}
\end{table}
we give the numerical values of the Kerr quasinormal frequencies corresponding to $l=2$ and $m=0$, obtained with the new technique. Note that, the values for the fundamental modes can be compared with Leaver's results and they match accurately, see Table $4$.
\begin{table}
\begin{tabular}{l|cc}
\hline
\hline
$\hat a$&Leaver's fund. mode&Our fund. mode\\ \hline
0.0&(0.747343, -0.177925)&(0.747343, -0.177925)\\
0.1&(0.750248, -0.177401)&(0.750252, -0.177401)\\
0.2&(0.759363, -0.175653)&(0.75936, -0.175653)\\
0.3&(0.766108, -0.171989)&(0.776104, -0.17199)\\
0.4&(0.803835, -0.164313)&(0.803834, -0.164314)\\
0.49&(0.844509, -0.147065)&(0.844509, -0.147065)\\
\hline
\hline
\end{tabular}
\caption{In this table, we compare our fundamental mode to Leaver's, for $l=2$ and $m=0$ and we show how close the two results are.}
\end{table}

\section{conclusion}
The quasinormal frequencies for the gravitational perturbation of a Kerr black hole are the most interesting from an astrophysical point of view and they have been matter of research for many years. 

Perturbations of rotating black holes can be reduced to the study of two continued fraction equations. Their common zeros are the angular and radial quasinormal frequencies. Searching for the solutions in a reliable way is a difficult task in numerical analysis. They could be found if we mapped out the full zero contours of both functions of the system. These zero contours would consist of an unknown number of disjoint closed curves. The roots are the intersections of pairs of zero contours \cite{flannery}.

In this paper we have introduced a new technique to compute such frequencies in a way that avoids the use of four dimensional root finding routines. The new technique consists of a new numerical way to evaluate the angular eigenvalues of the Teukolsky angular equation, which is independent from the solution of the radial part of the problem.

The results we have obtained work for any value of the spherical harmonic parameters $l$ and $m$ and for the angular parameter $a$ ranging from the static to the Kerr limit. These results accurately agree with former work and results are given for higher modes than hitherto. It would be interesting to analyse the possibility to extend the numerical computation to higher modes or to look for an analytical formula to describe the asymptotic behaviour of the Kerr quasinormal frequencies in a way similar to what has been already done in the static case \cite {nollert}.

It is intended that the {\em C++} programme used to obtain the results in this paper, will be made available to the worldwide community as a useful tool in gravitational wave research.

\end{document}